\documentclass{ws-p8-50x6-00}
\begin{document}

\def\relR{\mbox{\small \bf R}}
\def\inr{\mbox{\small \bf r}}
\def\tn{\tilde n}
\def\Rdot{\stackrel{\cdot }{\relR}}
\def\Rddot{\stackrel{\cdot \cdot }{\relR}}

\title{Effects of  shell structure in reactions
 leading to the same compound nucleus or different isotopes
\footnote{T\lowercase{alk given at the} I\lowercase{nternational}
C\lowercase{onference on} N\lowercase{uclear} P\lowercase{hysics at}
B\lowercase{order} L\lowercase{ines,}
NPBL2001, L\lowercase{ipari (}M\lowercase{essina,}
I\lowercase{taly), 21-24} M\lowercase{ay, 2001}}
}

\author{A.K. Nasirov$^{\dagger}$}
\address{Bogoliubov Laboratory of Theoretical Physics, JINR,
141980 Dubna, Russia\\
E-mail: nasirov@thsun1.jinr.ru\\}
\author{G. Fazio, G. Giardina, A. Lamberto, R. Ruggeri, A. Taccone}
\address{Istituto Nazionale di Fisica Nucleare,
 Sezione di Catania,  and\\
 Dipartimento di Fisica dell'Universit\`a di Messina, 98166 Messina,
Italy\\
E-mail: giardina@nucleo.unime.it}
\author{A.I. Muminov}
\address{$^{\dagger}$
Institute of Nuclear Physics 702132 Ulugbek, Tashkent, Uzbekistan\\
E-mail: ion@suninp.tashkent.su}
\author{F. Hanappe}
\address{Universit\'e Libre de Bruxelles, CP 229, B-1050 Bruxelles,
Belgium\\
E-mail: fhanappe@ulb.ac.be}
\author{R. Palamara}
\address{Facolt\`a di Architettura dell'Universit\`a
 di Reggio Calabria, Italy\\
E-mail: rpalamara@unirc.it}
\author{L. Stuttg\'e}
\address{Institut de Recherches Subatomiques, F-67037 Strasbourg Cedex 2,
France\\
E-mail: stuttge@in2p3.fr}


\maketitle

\abstracts{
The role of the entrance channel in fusion-fission
reactions was studied  by the  theoretical analysis of the
experimental   evaporation residue excitation functions for
reactions leading to the same compound nucleus. The evaporation
residues cross sections for $xn$-channels were calculated in the frame
of the combined dinuclear system concept (DNS) and advanced
statistical model (ASM). The revealed differences between
experimental data on the evaporation residues in the $^{40}$Ar+$^{176}$Hf,
$^{86}$Kr + $^{130}$Xe and $^{124}$Sn + $^{92}$Zr reactions leading
to the $^{216}$Th$^*$ compound nucleus (CN) are explained
by the different spin distributions of compound nuclei which are formed.
It is shown that the intrinsic fusion barrier $B^*_{fus}$ and size of
potential well are different for every entrance channel.
}

\section{Introduction}

An appropriate way to understand the mechanism of the fusion-fission
process in heavy ion collisions is the analysis and comparison of
the measured excitation functions of evaporation residues for
reactions leading to the same compound nucleus.
Theoretical calculations performed in the framework of the model
including the nuclear shell effect and shape of colliding nuclei
allow us to come to useful conclusions about the mechanism of
fusion-fission process.
Often excitation functions of evaporation residues measured in
various reactions leading to the same compound nucleus are 
different not only by the position of the maximum but also by
the value of their maximums. The question on what characteristics
of nuclei or what relevant degrees of freedom of the
fusion-fission process are responsible for such difference in
evaporation residue cross sections is a well actual problem to reveal
optimal conditions for the synthesis of new super-heavy elements.

\section{Evaporation residue production in the DNS concept}
\label{dnscon}

In the dinuclear system concept\cite{DNSV935} the
evaporation residue cross section is factorized as follows:
\begin{equation}
\label{evapor}
\sigma_{er}(E)=\sum_{\ell=0}^{\infty}(2\ell+1)\sigma_{\ell}^{fus}(E,
\ell)W_{sur}(E,\ell).
\end{equation}
Here the effects connected with the entrance channel  are included
by the partial fusion cross section $\sigma_{\ell}^{fus}(E)$ which
is defined by the expressions:
\begin{eqnarray}
\label{s_fus}
\sigma_{\ell}^{fus}(E)&=&\sigma_{\ell}^{capture}(E) P_{CN}(E,\ell),\\
\sigma_l^{capture}(E)&=&\frac{\lambda^2}{4\pi}{\cal
P}_{\ell}^{capture}(E),
\end{eqnarray}
where $\lambda$ is the de Broglie wavelength of the entrance
channel, $P_{CN}(E,\ell)$ is a factor taking into account the
decrease of the fusion probability due to break up of the
dinuclear system  before fusion; ${\cal P}_{\ell}^{capture}(E)$
is the capture probability which depends on the  collision
dynamics and determines the amount of partial   waves leading to
capture.

The number of the partial waves was obtained by   solving
the equation of motion for the relative distance and orbital
angular momentum
\begin{eqnarray}
\label{maineq}
&&\mu(R(t))\Rddot + \gamma_{R}(R(t))\Rdot(t)=
-\frac {\partial V(R(t))}{\partial R},\\
&&\frac{dL}{dt}=\gamma_{\theta}(R(t))\left(\dot{\theta} R_{eff}^2
-\dot{\theta_1} R_{1eff}^2
-\dot{\theta_2} R_{2eff}^2\right)\,,
\end{eqnarray}
where $R(t)$ is the relative motion coordinate, $\Rdot(t)$ is
the corresponding velocity; $\dot\theta$, $\dot\theta_1$ and
$\dot\theta_2$ are  angular velocities  of the dinuclear system and
its  fragments, respectively;  $\gamma_{R}$  and $ \gamma_{\theta}$
are the friction coefficients for the relative motion along $R$ and
tangential motion when two nuclei roll on each other's surfaces,
respectively;
$V(R)$ is the nucleus-nucleus potential; $\mu(R(t))$ is the reduced
mass of the system; $R_1$ and $R_2$  are the fragment radii.

Calculations showed that use of these kinetic coefficients leads to
gradual dissipation of kinetic and rotational energy  \cite{Adam97}.
It was shown that at collisions
of massive nuclei
despite of continuous dissipation the capture becomes
impossible at larger values of beam energy than the Coulomb barrier,
because of the small size of the well in the nucleus-nucleus potential.
The dissipation is not sufficient to trap colliding
nuclei in the potential well to create  a necessary condition for fusion,
and in this case the formed dinuclear system  undergoes quasi-fission.
The nucleus-nucleus potential  $V(R)$  depends on the
mutual orientations of the  symmetry axes of deformed nuclei relative to
$\relR(t)$. The quadrupole ($2^+$) and octupole ($3^-$) collective
excitations in spherical nuclei are taken into account.
Thus, it is possible to consider fusion at different initial
orientations of the  symmetry axes.

The competition between fusion and quasi-fission is taken into
account by the factor $P_{CN}(E,\ell)$ which is calculated using
the method developed  in \cite{DNSV935}, in framework of the
statistical model. It was suggested that the probability of
realizing the complete fusion (in competition with the quasi-fission,
starting from an entrance channel that gives a DNS), is related to the
ratio of level densities, depending on the
intrinsic fusion or quasi-fission barriers, by the expression:
\begin{equation}
\label{Pcn}
P_{CN}=\frac{\rho(E^*_{DNS} -
B^*_{fus})}{\rho(E^*_{DNS} - B^*_{fus}) + \rho(E^*_{DNS} -
B_{qf})},
\end{equation}
where $\rho(E^*_{DNS} - B^*_K)$ is the  level   density.

In Eq. (\ref{Pcn}),  $B_{qf}$ is the barrier of the
nucleus-nucleus interaction potential which needs to be overcome
if the dinuclear system  decays in two fragments, and $E^*_{DNS}$ is
the excitation energy of the dinuclear system  given by the
difference between  beam energy $E_{\rm c.m.}$  and the minimum of
the nucleus--nucleus potential ($E^*_{DNS}=E_{\rm c.m.}-V(R_m)$).

The advanced statistical model (ASM), described in detail in
\cite{ASM,Sag98}, allows us to take into account the dynamical
aspect of the fission-evaporation competition at  the compound
nucleus evolution along the de-excitation cascade.
Particular attention is
devoted to the determination of the level densities.

Dissipation effects, which delay fission, are treated according to
\cite{GraWeiPLB80,RastSJNP91}. These include Kramers' stationary
limit \cite{KramP40} and an exponential factor applied to Kramers'
fission width to account for the transient time, after which the
statistical regime is reached. The systematic obtained by
Bhattacharya {\it et al.}\cite{BhatPRC96} gives the possibility
of taking into account the incident energy per nucleon $\epsilon$
and compound nucleus mass $A_{cn}$ dependencies of the reduced
dissipation coefficient $\beta_{dis}$.

According to dinuclear system concept\cite{DNSV935,GiarSHE}, 
the fusion of two colliding nuclei is possible if the following  
two conditions are satisfied. The first one is  overcoming Coulomb 
barrier along axis connecting nuclear centers by nuclei at incoming 
stage of collision  and formation of the nuclear composite system
(molecular alike so called dinuclear system). This process is called a
capture. The second condition is the transformation of the dinuclear
system into the more
compact compound nucleus overcoming  the intrinsic barrier
($B^*_{fus}$) which must be overcome by system during the motion
on the mass (charge) asymmetry axis. So this stage is a fusion and
it is realized by nucleon transfer from the light nucleus on the
unoccupied states of heavy one at the limited size of overlap
volume of nucleon density. For light and mediate nuclear system or
for heavy nuclear system with largest mass asymmetry this $B^*_{fus}$
barrier is very small or equal to zero and capture leads immediately
to fusion. Therefore for those cases fusion cross section is calculated
in frame of well known models.
\begin{figure}[t]
\epsfxsize=20pc 
\centerline{\epsfbox{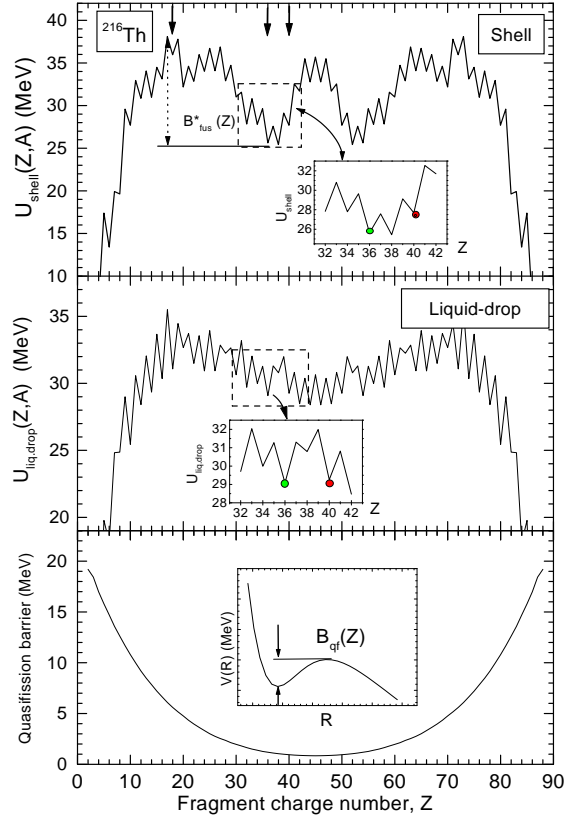}} 
\vspace*{-1.0cm}
\caption{Driving potential $U(Z,A,R_m;l=0)$ as
function of the charge number  $Z$ of a fragment of dinuclear
system calculated by using of binding energies from the mass table$^9$
(top panel) and liquid-drop model (middle-panel).
The vertical arrows indicate
initial charge number of light nuclei in the  $^{40}$Ar+$^{176}$Hf (I), 
 $^{86}$Kr + $^{130}$Xe (II) and $^{124}$Sn + $^{92}$Zr (III) reactions
leading to $^{216}$Th. The intrinsic  $B^*_{fus}$ (top panel) and 
quasifission  $B_{qf}$ (bottom panel) barriers are shown as a function of 
charge number of a fragment of dinuclear system.} 
\label{fig1}
\end{figure}
 The  barrier $B^*_{fus}$ is determined by the difference between
the maximum value of a driving potential  and its value at the
point corresponding to the initial charge asymmetry of the
considered reaction (Fig.\ref{fig1}). A driving potential holds
significance in consideration of  the fusion process as a motion
of system along  the charge (mass) asymmetry degree of freedom.
Therefore, here more attention should be paid to that. It is
obtained from the potential energy surface $U(A,Z;R,\ell)$ as a
function of masses (charges) $A_1, A_2$ ($A_2=A_{tot}-A_1 $) of
fragments forming the dinuclear system at the $R_m$ value of
internuclear distance corresponding to the minimum of their
nucleus-nucleus potential $V(R)$\cite{GiarSHE}.
 The driving potential
$U(A,Z;\ell)$ calculated in this way is presented in
Fig.\ref{fig1}. The distribution of neutrons between two
fragments by the given proton numbers  $Z_1$ and $Z_2$ (or ratios
$A_1/Z_1$ and $A_2/Z_2$ for the both of fragments) was determined
by minimizing of potential $U(A_1,Z_1;R)$ as a function of $A_1$
for each $Z_1$.
\begin{equation} \label{PES}
U(A,Z;R_m,\ell)=B_1+B_2+V_0(Z,\ell,\beta_1,\alpha_1;
\beta_2,\alpha_2; R_m)-(B_0+V_{CN}(\ell)),
\end{equation}
where, $B_1$,  $B_2$ and $B_0$ are the binding energies of the
nuclei forming a dinuclear system and of the compound nucleus,
respectively. The values of $B_i (i=0,1,2)$ were obtained from\cite{MassAW95};
$\beta_i (i=1,2)$ are  the fragment deformation
parameters and $\alpha_i$ are the orientations of nuclei symmetry
axis relative to the beam direction; $V_{CN}(\ell)$ is
the rotational energy of the compound nucleus. The $R_m$ is the
position of the minimum of the nucleus-nucleus potential (bottom of the pocket)
on the $R$ axis for a given mass asymmetry $A_1$. The smallest value of
excitation energy of the compound nucleus ($E^{*(min)}_{CN}$) is determined
by the absolute maximum value of the driving potential laying on
the way to fusion ($Z=0$)  from the point  corresponding to the
initial charge asymmetry (Fig.\ref{fig1})
for the given mutual orientations of the axial symmetry axes of 
the projectile- and target-nucleus. Because the shapes of the potential 
energy surface and driving
potential depend on the orientations of nuclei relatively to axis
connecting the centers of interacting nuclei. The presented
results on capture and fusion cross sections are obtained by averaging 
over contributions of different
orientations. The quasi-fission is a decay of dinuclear system
 moving along $R$  in the $V(R)$ nucleus-nucleus
interaction potential without reaching compact shape. Thus, for
quasi-fission, it is necessary to overcome the Coulomb barrier on
the $R$ axis from internal side. The DNS concept allows one to reveal a
reason of the strong decrease in the fusion cross section for a massive system
or for a symmetric entrance channel due to the increase of $B^*_{fus}$.

\section{Comparison of DNS model results and experimental data}
\subsection{Reactions leading to $^{216}$Th}
\label{dnsexp}
 The role of the entrance channel on the formation
of the compound nucleus and evaporation residues is a main point of our
interest. The qualitative difference  between  fusion excitation
functions of reactions leading to the same compound nucleus allows
us to analyze  the effect of the shell structure  on the fusion
mechanism. Experimental excitation functions of the evaporation
residues  measured in  $^{86}$Kr + $^{130, 136}$Xe \cite{Ogan96}
reactions in the Flerov Laboratory (Dubna) and that for the
$^{40}$Ar+$^{176}$Hf \cite{Verm84,Clerc84} and $^{124}$Sn +
$^{92}$Zr \cite{Sahm85} reactions have been compared with the
results of calculation in the frame of the method presented 
in\cite{GiarSHE}.
It is shown that the effect of shell structure is
revealed in the differences at comparison of spin distributions of
compound nucleus formed by using different reactions.

\begin{figure}[t]
\epsfxsize=20pc 
\centerline{\epsfbox{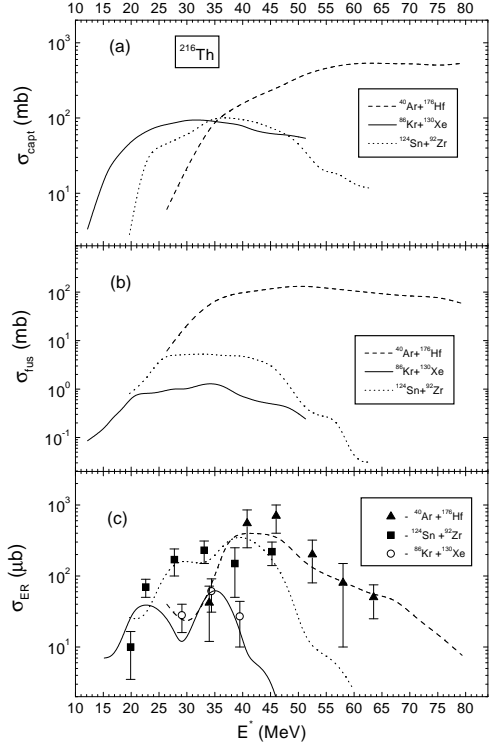}} 
\vspace{-2.2cm} \caption{Comparison of the calculated capture (a) and
fusion excitation function (b) as well as calculated total excitation
functions of evaporation residue  with the experimental data (c) for
$^{40}$Ar+$^{176}$Hf$^{12}$, $^{124}$Sn + $^{92}$Zr$^{14}$, and $^{86}$Kr +
$^{130}$Xe$^{11}$  combinations leading to the $^{216}$Th CN.}
\label{fig2}
\end{figure}

Experimental data reveal that the maximum value of the
evaporation residue for $^{40}$Ar+$^{176}$Hf (I) is twelve   times 
larger than for
$^{86}$Kr + $^{130}$Xe (II) and three  times  larger than for
$^{124}$Sn + $^{92}$Zr (III) (see Fig.\ref{fig2}). Because the
$^{40}$Ar+$^{176}$Hf reaction has largest charge asymmetry
($\eta_{Z}=(Z_{2}-Z_{1})/(Z_{1}+Z_{2})$) in comparison with two
others (II,III) and the height of barrier $B^*_{fus}$ in the way
to fusion for this reaction is smaller than that for (II,III) (Table
\ref{tabth216}). 
The way to fusion is longer for the dinuclear system which has more
mass symmetric configuration. In Figs.\ref{fig2}a and \ref{fig2}b, 
the excitation functions of the  capture and fusion calculated in the 
frame of  dinuclear system for the (I,II,III) reactions are compared. 
The excitation functions of the evaporation residue calculated in the frame 
of advanced statistical model\cite{ASM} are in good agreement with the
experimental data (see Fig.\ref{fig2}c).
At these calculations, spin distributions for
CN were found by the method presented in\cite{GiarSHE}.

\begin{table}[t]    
\caption{Charge asymmetry, intrinsic fusion $B^*_{fus}$,
quasi-fission $B_{qf}$ barriers and the hindrance factor $P_{CN}$
for the reactions leading to
$^{216}$Th$^*$.}
\label{tabth216}
\begin{center}
\footnotesize
\begin{tabular}{|c|c|c|c|c|}
\hline Reactions &\raisebox{0pt}[13pt][7pt]{$\eta_{Z}$} &
\raisebox{0pt}[10pt][7pt]{$B^*_{fus}$ (MeV)}
&\raisebox{0pt}[10pt][7pt]{$B_{qf}$ (MeV)}&\raisebox{0pt}
[10pt][7pt]{$P_{CN}$}\\\hline
\raisebox{0pt}[10pt][7pt] (I) \quad\, {$^{40}$Ar+$^{176}$Hf}  & 0.60 &
2.31 &
7.12 & 0.580 \\
\raisebox{0pt}[10pt][7pt] (II) \, {$^{86}$Kr + $^{130}$Xe}  &0.20 &
12.31&
2.35 & 0.014 \\
\raisebox{0pt}[10pt][7pt] (III) \, {$^{124}$Sn + $^{92}$Zr} & 0.15& 9.87
&
1.35 & 0.081 \\
[4pt]
\hline
\end{tabular}
\end{center}
\end{table}

But an unusual phenomenon is that the maximum value of the evaporation 
residue for
$^{124}$Sn + $^{92}$Zr  is four times larger than for $^{86}$Kr +
$^{130}$Xe nearly at the same value of $E^*_{CN}$. These reactions
lead to the same ($^{216}$Th$^*$) CN. The mass asymmetry of (III)
$((A_2-A_1)/(A_1+A_2)$=0.148) is smaller than one of (II)
(0.203).  It is not clear from the driving potential calculated
using  the binding energies $B_1, B_2$
and $B_{CN}$ determined by the liquid-drop model (see middle
panel of Fig.\ref{fig1}), since the value of the fusion barrier
$B^*_{fus}$ for the $^{124}$Sn + $^{92}$Zr reaction is not lower
than that for the $^{86}$Kr + $^{130}$Xe reaction.

The reason of this phenomenon can be account for by comparison
of driving potentials calculated using binding energies  obtained 
from the mass table \cite{MassAW95} and that  determined 
by liquid drop model (see Fig.\ref{fig1}, top and middle
panels, respectively). The values of driving potential corresponding 
to the $^{86}$Kr + $^{130}$Xe and  $^{124}$Sn + $^{92}$Zr reactions
in the top and middle panel are different.
One can see that, in the top panel of Fig.\ref{fig1},
intrinsic fusion barriers for $B^*_{fus}$ for the 
$^{86}$Kr + $^{130}$Xe reaction is larger than that 
$^{124}$Sn + $^{92}$Zr reaction. In this case, the hindrance factor 
$P_{CN}$ of the fusion cross cross is larger for the  former  reaction 
than last one.
This corresponds to the observed phenomenon for these two reactions.  
This discloses even qualitatively difference between
values of $B^*_{fus}$ for reactions (II) and (III). 
It is seen from the middle panel of Fig.\ref{fig1}
that $B^*_{fus}(II) \approx B^*_{fus}(III)$ when the driving 
potential is calculated using the binding energies
 $B_1, B_2$ and $B_{CN}$ using the 
liquid-drop model. If the last case takes place then 
a difference between the  fusion excitation function 
of the $^{86}$Kr + $^{130}$Xe and 
$^{124}$Sn + $^{92}$Zr reactions  must be very small. 
This is in contradiction with experimental data and 
that means using of binding energies obtained in the 
liquid-drop model is not suitable to analyze such 
unusual phenomenon. As seen from
the Table \ref{tabth216} for the former reaction $B^*_{fus}$ is
larger than for the latter one due to shell effects for nuclear
binding energy in the range of charge number of light fragment 
$Z=30\div40$.

Another reason is seen from the analysis
of spin distributions of compound nuclei formed in these three
reactions.
\begin{figure}[t]
\epsfxsize=20pc 
\centerline{\epsfbox{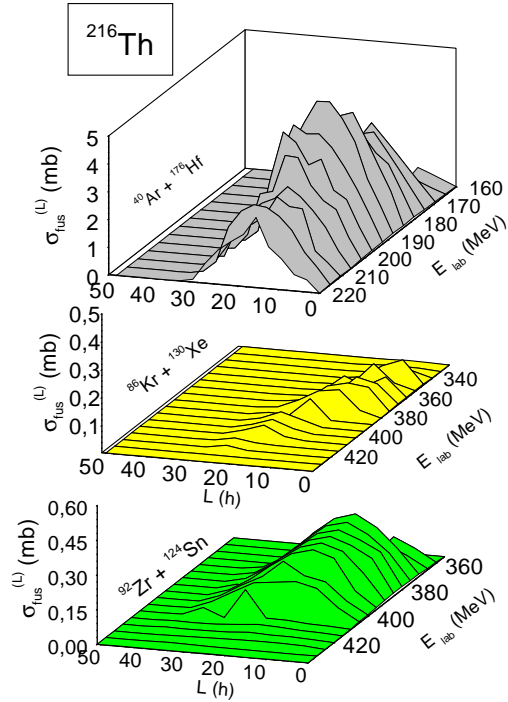}} 
\vspace*{-2.8cm}
\caption{Spin distributions for the
$^{40}$Ar+$^{176}$Hf (top panel), $^{86}$Kr + $^{130}$Xe (middle)
and $^{124}$Sn + $^{92}$Zr (bottom) reactions.}
\label{fig3}
\end{figure}

As seen from  Fig.\ref{fig3}, spin distribution of CN formed in
reaction (I) (top panel) has largest volume in comparison for the
reactions (II) (middle panel) and (III) (bottom panel). It is an
expected result. The matter is  that why the volume of in spin
distributions of CN corresponding to  reaction (III) is larger
than that for reaction (II). This could be connected with the
$\sigma_l^{capture}(E)$ and $P_{CN}$  which determine partial
fusion cross sections (\ref{s_fus}). The last is calculated
from the equation of motion. The number of partial waves
$\ell$  which contribute to $\sigma_l^{capture}(E)$ is determined
by the size of the potential well in the entrance channel (the high
of quasi-fission barrier $B_{qf}$) and the friction coefficient
of the radial motion. The use of the calculated
friction coefficients leads to a gradual dissipation of kinetic and
rotational energies\cite{Adam97}. It was shown that by collisions
of massive nuclei, despite of continuous dissipation, the capture
becomes impossible at beam energy values larger than the
Coulomb barrier because of the small size of the well in the
nucleus-nucleus potential which is a function of $R$. The
dissipation is not enough to trap colliding nuclei in the
potential well to create a necessary condition for fusion at low
values of angular momentum which allow to the dinuclear system to fuse. At
largest values of beam energy, capture is possible only for high
angular momentum.  In this case the formed dinuclear system can
exist as a molecular state forming a super-deformed shape or
undergoes quasi-fission because $B_{fus}^*$ increases by
angular momentum of dinuclear system. Then the maximum value of driving 
potential increases more faster than its value corresponding to 
charge (mass) asymmetry of the entrance channel. Therefore the maximum 
of the calculated spin distributions has tendency to move to larger
values of angular momentum at beam energies well above the
Coulomb barrier.

This is seen in the spin distributions for the
$^{124}$Sn + $^{92}$Zr  and $^{86}$Kr + $^{130}$Xe reactions
(Fig.\ref{fig3}). The driving potential is the same for all the
reactions leading to the same compound nucleus. Therefore the
intrinsic fusion barriers for these reactions under discussion can
be compared. It is seen in Fig.\ref{fig1} (top panel) from the curve 
of the driving potential that $B^*_{fus}$ is smaller for the reaction
(III) than for (II). This is connected with the increase of the
potential energy at $Z>37$ due to the shell effects.

The calculation of the driving potential using binding
energies determined by liquid drop model shows that in this case
the shape of the curve is different than one calculated using
values of binding energies from the mass table in \cite{MassAW95}.
Therefore, an observed smallness of the excitation function of
evaporation residues for the $^{86}$Kr + $^{130}$Xe reaction than that for
$^{124}$Sn +$^{92}$Zr is concluded to be connected with the peculiarity of
the shell structure of nuclear fragments forming the dinuclear system.
Due to large difference between $Q$-values of these three reactions
leading to the $^{216}$Th$^*$ CN, the centers of their excitation
functions are placed at different values of the excitation energy.

\subsection{Comparison of reactions induced by  $^{86}$Kr on the
$^{130}$Xe and $^{136}$Xe targets}

\begin{figure}[t]   
\epsfxsize=18pc 
\centerline{\epsfbox{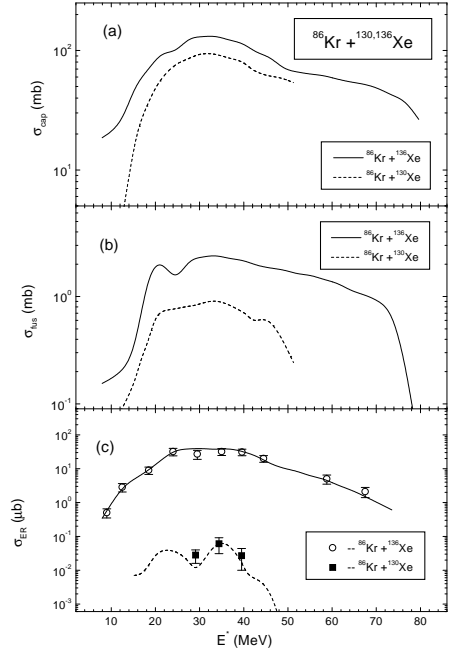}} 
\vspace*{-1.7cm}
\caption{Comparison of the calculated capture (a),
fusion (b) and evaporation residue (c) excitation functions as
well as the measured  excitation functions of evaporation residue
(c) for the $^{86}$Kr + $^{136}$Xe$^{11}$ (solid curve, open circles)
and  $^{86}$Kr + $^{130}$Xe$^{11}$ (dashed curve, solid squares)
reactions.}
\label{fig4}
\end{figure}

\begin{figure}[t]  
\epsfxsize=18pc 
\centerline{\epsfbox{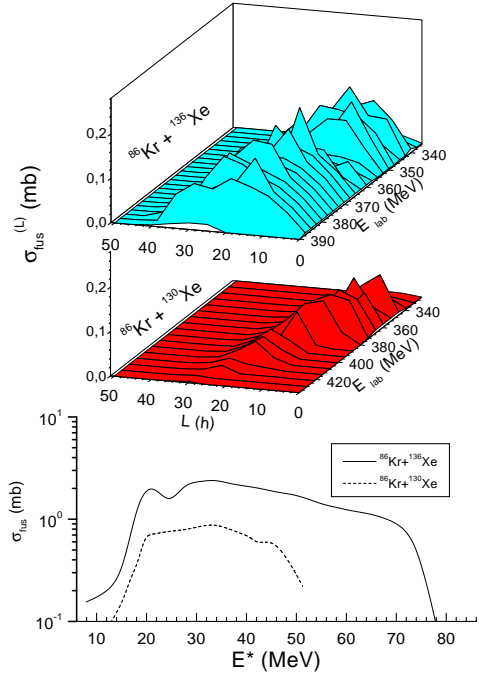}} 
\vspace*{-1cm}
\caption{Spin distribution for the
$^{86}$Kr+$^{136}$Xe (top panel) and  $^{86}$Kr + $^{130}$Xe
(middle panel) reactions at different beam energies  $E_{lab}$.
Comparison of fusion excitation functions for these reactions 
(bottom panel).}
\label{fig5}
\end{figure}

Another observed interesting phenomenon was revealed at
the comparison of the experimental data on reactions induced by
the $^{86}$Kr projectile on the $^{130}$Xe and $^{136}$Xe targets.
The evaporation residue in $^{86}$Kr + $^{136}$Xe (IV) was about 500 times 
larger than in $^{86}$Kr + $^{130}$Xe (II) (Fig.\ref{fig4}c). This
result is related to the two characteristics of the fusion-fission
mechanism. At first,  the fusion cross section calculated using
the model reviewed in \cite{GiarSHE}
for the reaction $^{86}$Kr + $^{136}$Xe is well larger than that for
$^{86}$Kr + $^{130}$Xe (Fig.\ref{fig4}b). It is seen from
the difference in spin distributions presented in
Fig.\ref{fig5} for these two reactions. 

\begin{table}[t]    
\caption{Charge asymmetry, intrinsic fusion $B^*_{fus}$,
quasi-fission $B_{qf}$ barriers and the hindrance factor $P_{CN}$
 for the reaction leading to $^{222}$Th$^*$.}
\label{tabth222}
\begin{center}
\footnotesize
\begin{tabular}{|c|c|c|c|c|}
\hline Reactions &\raisebox{0pt}[13pt][7pt]{$\eta_{Z}$} &
\raisebox{0pt}[10pt][7pt]{$B^*_{fus}$ (MeV)}
&\raisebox{0pt}[10pt][7pt]{$B_{qf}$ (MeV)}&\raisebox{0pt}
[10pt][7pt]{$P_{CN}$}\\\hline
\raisebox{0pt}[10pt][7pt] (II) \, {$^{86}$Kr + $^{130}$Xe}  &0.20 &
12.31&
2.35 & 0.014 \\
\raisebox{0pt}[10pt][7pt]
(IV) \quad\, {$^{86}$Kr + $^{136}$Xe} & 0.20 & 7.52 & 4.05 & 0.027 \\
[4pt]
\hline
\end{tabular}
\end{center}
\end{table}

This  difference could  be explained by the two facts:  the size
of the well in the nucleus-nucleus potential for the $^{86}$Kr +
$^{130}$Xe (II) reaction is smaller than that for $^{86}$Kr +
$^{136}$Xe (IV) and the intrinsic fusion barrier 
is for the (II) reaction ($B^*_{fus}$=12.3 MeV) larger than for the (IV)
(7.5 MeV) while the quasifission barrier $B_{qf}$ is 2.3 MeV for the DNS
obtained by $^{86}$Kr+$^{130}$Xe and 4.05 MeV for $^{86}$Kr +
$^{136}$Xe (see Table \ref{tabth222}).

Therefore, by comparing these two reactions, a higher
value of $B_{qf}$ (4.05 MeV against 2.3 MeV) together with a lower
value of $B^*_{fus}$ (7.5 MeV against 12.3 MeV) lead to a higher
value of the fusion cross section for the reaction that gives the
$^{222}$Th* CN in comparison with the reaction that gives the
$^{216}$Th* CN.
Because the excess number of neutrons in
$^{136}$Xe makes the nuclear interaction more attractive and the
potential well is more wide in comparison with the nucleus-nucleus
potential between $^{86}$Kr and $^{130}$Xe. The calculated
excitation functions for the sum of the evaporation residues
in the $^{86}$Kr + $^{130}$Xe and $^{86}$Kr + $^{136}$Xe
reactions are shown in Fig.\ref{fig4}c.

At second,  the
survival probability ($W_{sur}$) of $^{222}$Th$^*$ CN is larger
than one of $^{216}$Th*: fission barrier of the isotope $A=222$  is
larger than for the isotope $A=216$ while the neutron separation energy
of $^{222}$Th* ($S_n=7.808$ MeV) is smaller than $^{216}$Th*
($S_n=8.701$ MeV). Therefore the $\Gamma_n/\Gamma_f$ ratio is larger for
$^{222}$Th$^*$  than for $^{216}$Th$^*$. For these two compound nuclei
the $\Gamma_n/\Gamma_f$ ratio is very different at each step of
the two  de-excitation cascades and it is larger for the cascade
of $^{222}$Th*.  By comparing the $\Gamma_n/\Gamma_f$ values at
each step of the cascades of $^{222}$Th* and $^{216}$Th*, starting
from the same excitation energy of the compound nuclei, in the
30-42.5 MeV excitation energy range of CN, we find  the values of the ratio
$(\Gamma_n/\Gamma_f)_{^{222}Th*} /
(\Gamma_n/\Gamma_f)_{^{216}Th*}$ to be between 1.3 and 4.2 for the
1$n$-channel; this ratio  is included between 2.2 and 9.9 for the
2$n$ channel; its values  are included between 3.0 and $1.9\times
10^5$ for the 3$n$ channel. The above-mentioned ratio is between
$2.7\times 10^4$ and $9.1\times 10^5$ for the 4$n$ channel, in the
36.5-42.5 MeV excitation energy range of the compound nuclei.
Large $\Gamma_n/\Gamma_f$ values correspond to a large evaporation
residue production.

\section{Conclusion}

The combined dynamical and statistical model based on the dinuclear system
approach allowed us to estimate excitation functions of quasi-fission,  fusion,
and formation of evaporation residues in  fusion  reactions with
massive nuclei. The capture stage was calculated using the dynamical model
and for calculation of the fusion stage a statistical   approach was used.
The obtained optimal beam energy or excitation energy of the compound
nucleus is in good agreement with the
experimental data. The  fusion excitation functions calculated in this way
 were used to estimate the surviving probability of the formed compound
nucleus relative to fission in the frame of the advanced statistical model
for the de-excitation cascade. The excitation functions  obtained in this
paper are in good agreement with the ones measured in experiments.

The effect of the entrance channel was studied by analyzing the quantities
in detail which were used
in the calculation of the fusion cross section. These are the capture cross
section (which is the formation probability of the dinuclear system in
competition with quasi-fission), the intrinsic fusion barrier $B^*_{fus}$,
 the quasi-fission barrier $B_{qf}$, and the  excitation energy $E^*_{DNS}$
of the dinuclear system.

According to the scenario of the DNS-concept, the comparison of the data
leading to the different $^{216}$Th$^*$ and $^{222}$Th$^*$ isotopes
reveals a great role of the dynamics of the entrance
channel and the  nuclear shell structure on the mechanism of the
complete fusion and the evaporation residue formation.

\vspace*{-0.4cm}

\section*{Acknowledgments}
\vspace*{-0.3cm}
We are grateful to Prof. R.V. Jolos, Prof. V.V. Volkov, Prof. Dr.
W. Scheid, Drs. G.G. Adamian, and N.V. Antonenko for helpful
discussions. One of the authors (A.K.N.) thanks the INTAS 
(Grant No. 991-1344), the
Russian Fund of Basic Research (Grant No. 01-02-16033)  for financial
support, and he is grateful to the STCU Uzb-45,
Uzbekistan State Scientific-Technical Committee (Grant No. 7/2000)
and   Fund of Uzbek Academy of Science for Support of Basic Research 
N45-00 for partial support. A.K.N. would also like to express his gratitude
to the Universit{\`a} of Messina (Italy) for the warm hospitality during
his stay, and to the Fondazione Bonino-Pulejo (FBP) of Messina for the
help given in developing the collaboration with the group of Prof. G.
Giardina.


\vspace*{-0.4cm}

\begin{thebibliography}{99}
\vspace*{-0.2cm}
 \bibitem{DNSV935} 
 V.V. Volkov, N.A. Antonenko, E.A. Cherepanov, A.K. Na\-si\-rov, 
V.P. Permjakov,
 \Journal{\PLB} {319} {425} {1993}; \Journal{\PRC} {51} {2635} {1995}.

\bibitem{Adam97} 
 G.G.~Adamian, R.V.~Jolos, A.I.~Muminov,  A.K.~Nasirov,
 \Journal{\PRC} {56} {373} {1997}.

\bibitem{ASM} 
 A.~D'Arrigo, G.~Giardina, M.~Herman, A.V.~Ignatyuk, A.~Taccone, J. Phys. G
 {\bf 20}, 365 (1994).

\bibitem{Sag98} 
 R.N.~Sagaidak, V.I~Chepigin, A.P.~Kabachenko, J.~Roh\'a\v{c},
 Yu.Ts.~Oganessian, A.G.~Popeko, A.V.~Yeremin, A.~D'Arrigo, G.~Fazio,
 G.~Giardina, M.~Herman, R.~Ruggeri, and R.~Sturiale, J. Phys. G {\bf
 24}, 611 (1998).

\bibitem{GraWeiPLB80} 
 P. Grange and H.A. Weidenm{\"u}ller, Phys. Lett.
{\bf 96B}, (1980) 26.

\bibitem{RastSJNP91} 
E.M. Rastopchin, S.I. Mulgin, U.V. Ostapenko, V.V. Pashkeevich,
M.I. Svirin, G.N. Smirenkin, Sov. J. Nucl. Phys. {\bf 53},  (1991)
741.

\bibitem{KramP40} 
H.A. Kramers, Physica {\bf 7},  (1940) 284.

\bibitem{BhatPRC96} 
 C. Bhattacharya, S. Bhattacharya, and K. Krishan,
 Phys. Rev.  {\bf C 53},  (1996) 1012.

\bibitem{GiarSHE} 
G. Giardina, S. Hofmann, A.I. Muminov, A.K. Nasirov,
\Journal{\EPJA} {8}  {205} {2000}.


\bibitem{MassAW95} 
 G. Audi, A.H. Wapstra, \Journal{\NPA} {595} {509} {1995}.

\bibitem{Ogan96} 
 Yu.Ts.~Oganessian, A.Yu.~Lavrentev, A.G.~Popeko, R.N.~Sa\-gai\-dak, 
A.V.~Yeremin,
 S.~Hofmann, F.P. He{\ss}\-ber\-ger, V.Ninov, Ch.~Stodel, {\it JINR FLNR
 Scientific Report 1995-1996. Heavy Ion Physics}, B.I.~Pustylnik {ed.}, p.~62
 (JINR, E7-97-206, Dubna), {1997}.

\bibitem{Verm84} 
 D.~Vermeulen, H.-G.~Clerc, C.-C.~Sahm, K.-H.~Schmidt, J.G.~Keller,
 G.~M{\"u}nzenberg, W.~Reisdorf, \Journal{\ZPA} {318} {157} {1984}.

\bibitem{Clerc84} 
 H.-G.~Clerc, J.G.~Keller, C.-C.~Sahm, K.-H.~Schmidt, H.~Schulte, D.~Vermeulen,
 \Journal{\NPA} {419} {571} {1984}.

\bibitem{Sahm85} 
 C.-C.~Sahm, H.-G.~Clerc, K.-H.~Schmidt, W.~Reisdorf, P.~Armbruster,
 F.P.~He{\ss}berger, J.G.~Keller, G.~M{\"u}nzen\-berg, D.~Vermeulen,
 \Journal{\NPA} {441} {316} {1985}.


\end{thebibliography}
\end{document}